\documentclass[10pt, onecolumn, conference]{IEEEtran}

\usepackage[T1]{fontenc}
\usepackage[utf8]{inputenc}
\usepackage[english]{babel}
\usepackage{array}
\usepackage{graphicx}
\usepackage{booktabs}
\usepackage{amsmath}
\usepackage{amssymb}
\usepackage{multirow}
\usepackage{caption}
\usepackage{xcolor}
\usepackage{url}
\usepackage{float}  
\usepackage{breakcites}
\usepackage[hidelinks]{hyperref}
\usepackage{orcidlink}

\providecommand{\doi}[1]{\url{https://doi.org/#1}}

\graphicspath{{figures/}}

\makeatletter\g@addto@macro\UrlBreaks{\do\-\do\/\do\.\do\_}\makeatother
\makeatletter
\renewcommand\paragraph{\@startsection{paragraph}{4}{\z@}%
  {0.7ex \@plus.2ex \@minus.2ex}{-0.6em}%
  {\normalfont\normalsize\bfseries}}
\makeatother

\makeatletter
\renewcommand\@makefntext[1]{\noindent\scriptsize\@thefnmark.\ #1}
\makeatother
\usepackage[para]{footmisc}

\renewcommand{\arraystretch}{0.95}
\begin{document}

\title{Not All 4-bit Quantizers Are Equal:\\
Deployment-Time Mitigation of PII Leakage\\
in Fine-Tuned Small Language Models}

\author{
\IEEEauthorblockN{Cristhian Kapelinski\orcidlink{0009-0005-5750-022X}\IEEEauthorrefmark{1},
Diego Kreutz\orcidlink{0000-0003-0830-0238}\IEEEauthorrefmark{1}}
\IEEEauthorblockA{\IEEEauthorrefmark{1}AI Horizon Labs, Federal University of Pampa (UNIPAMPA), Alegrete, Brazil}
\IEEEauthorblockA{cristhianavila.aluno@unipampa.edu.br, diegokreutz@unipampa.edu.br}
}

\maketitle

\begin{abstract}
Organizations fine-tune small language models on private data and then compress them to 4 bits for resource-efficient deployment. We show that the compression method also affects privacy. What separates the methods is not the bit width but whether they tune their rounding on a small sample of text, the calibration corpus. On our primary model, when each planted record's own opening text is used as the prompt, the two calibration-based methods we test, Activation-aware Weight Quantization (AWQ) and Gradient-based Post-Training Quantization (GPTQ), each reproduce none of the planted records, while the calibration-corpus-free GGUF Q4\_K\_M format reproduces 5.3\% of them. Tracked across five open models with 0.5--7 billion parameters, AWQ leaks least at every size and in both families, with little accuracy loss at 3--7 billion. Controlled experiments associate the difference with calibration-induced rounding error in channels involved in rare-token prediction. Choosing the 4-bit method is therefore a deployment-time privacy decision, not only a question of speed and quality.
\end{abstract}


\section{Introduction}

Small language models (SLMs) with 1--7 billion parameters support on-premise deployment when cost, latency, or data-governance requirements limit the use of cloud services~\cite{husom2025edge}. Analysts project that artificial-intelligence-capable personal computers will account for 54.7\% of worldwide shipments by 2026.\footnote{\href{https://www.computerworld.com/article/4047019/ai-pcs-to-surge-claiming-over-half-the-market-by-2026.html}{Gartner via Computerworld, 2025}} Open-weight model families also give an organization a practical starting point for domain-specific fine-tuning, because adapting a model of this size no longer requires a datacenter~\cite{dettmers2023qlora,wang2025lora}. This deployment setting keeps prompts and inference inside the organization, but it does not remove the information the model memorized during training~\cite{carlini2023,lukas2023}.

A common deployment pipeline has two stages~\cite{zhang2025,lin2024awq}. First, an organization fine-tunes a model on internal data that may contain personally identifiable information (PII), such as email, customer records, or clinical notes. Second, it applies 4-bit post-training quantization (PTQ) to reduce the model's memory footprint and hardware requirements. A model's weights are ordinary numbers, normally stored in 16 bits each. Quantization stores each in 4 bits instead: weights are handled in small blocks, and within a block each becomes one of only 16 levels, plus one number per block saying how far apart those levels are. The model shrinks by roughly the ratio of the two widths and runs faster, since less data moves to the processor; the cost is that every weight is now slightly off. On Llama-2-7B that cost is small: 4-bit AWQ moves perplexity from 5.47 to 5.60, about 2\%, where perplexity measures how surprised the model is by text it never saw and \emph{lower} is better, so 5.60 is a slightly worse model. In exchange it generates tokens $3.2$--$3.3\times$ faster than the 16-bit version and runs a 13-billion-parameter model on an 8\,GB laptop GPU that cannot hold even a 7-billion-parameter model at 16 bits~\cite{lin2024awq}. Four bits is not only a small-deployment choice: OpenAI's open-weight gpt-oss models ship their largest weight matrices in a 4-bit floating-point format, which is what lets the 120-billion-parameter version run on one 80\,GB GPU,\footnote{\href{https://huggingface.co/openai/gpt-oss-120b}{OpenAI, gpt-oss-120b model card, 2025}} and Kimi K2 Thinking is served in native 4-bit integer precision, with every published benchmark number measured at that precision.\footnote{\href{https://huggingface.co/moonshotai/Kimi-K2-Thinking}{Moonshot AI, Kimi K2 Thinking model card, 2025}}

The 4-bit methods in common use disagree on where to put the rounding boundaries, and that disagreement is what we study. GGUF k-quants, the 4-bit formats shipped with the \texttt{llama.cpp} runtime and represented here by Q4\_K\_M, look only at the weights: each block is rescaled so its own largest weight fits the 4-bit range, and every block is treated alike~\cite{kurt2026llamacpp,husom2025edge}. How the model is actually used never enters the decision, so no extra data is needed. Activation-aware Weight Quantization with 4-bit integers (AWQ-4bit) instead first runs a small sample of text, the \emph{calibration corpus}, through the model, records how strongly each input channel responds, and scales up the channels that respond most before rounding~\cite{lin2024awq}. Rounding error is thereby pushed off the channels the calibration text exercises and onto those it leaves idle. We ask whether that redistribution also changes how much memorized PII the deployed model hands back. The question is first of all one of data security: a model fine-tuned on internal email or customer records is itself a copy of that data, and any user who can query it is a path out of the organization that never touches the source database. It is also a compliance question, since reproducing a person's data on demand bears on data-subject rights under the Brazilian General Data Protection Law (LGPD, Art.~18)\footnote{\href{https://www.planalto.gov.br/ccivil_03/_ato2015-2018/2018/lei/l13709.htm}{Lei n.\ 13.709/2018 (LGPD), Art.~18}} and the European General Data Protection Regulation (GDPR, Arts.~15 and~17).\footnote{\href{https://eur-lex.europa.eu/eli/reg/2016/679/oj/eng}{Regulation (EU) 2016/679 (GDPR), Arts.~15 and~17}}

Language models can reproduce training records to users who have only black-box query access. The standard way to measure this is to plant records in the training data on purpose, so that their later reproduction is proof of memorization rather than coincidence; such planted records are called \emph{canaries}~\cite{carlini2019,carlini2021}. Memorization grows with model capacity, duplication, and prompt length~\cite{carlini2023}, and PII leaks after fine-tuning too~\cite{lukas2023}. How the model was fine-tuned shifts the risk: updating only the last layer, the one that turns the model's internal state into a score per word, can leak more than updating the whole network~\cite{mireshghallah2022}; Low-Rank Adaptation (LoRA), which freezes the original weights and trains a small extra matrix beside each of them, can leak less than updating everything~\cite{wang2025lora}; and divergence attacks, which push the model off its normal behaviour with a degenerate prompt such as one word repeated indefinitely, make it fall back on reciting training text~\cite{nasr2023scalable}.

Two questions can be asked of a deployed model, and they are not equally severe. \emph{Verbatim extraction} asks whether the model will type out a memorized record when prompted; it needs only the ability to send text and read the answer, and a success is the record itself, in the clear. \emph{Membership inference} asks only whether a record was in the training set; it never recovers content, and it needs more: an interface that returns the probability the model assigns to each token of a candidate sequence. Neither needs the weights. We therefore treat verbatim extraction as the primary threat, since it both discloses data and needs the weaker interface, and report membership inference as a complementary measure.

At a fixed 4-bit budget, does the quantization method change the extraction of PII memorized during fine-tuning? The closest prior study compares round-to-nearest (RTN), AWQ, and Gradient-based Post-Training Quantization (GPTQ) on a 7-billion-parameter model after machine unlearning, a procedure that edits a trained model to remove specific records, and finds no method-specific difference~\cite{zhang2025}; related studies explain that coarse quantization can erase the small weight update unlearning produces~\cite{mishra2026quail,kupssinsku2026loraunlearn}. We instead quantize immediately after a fresh fine-tune, whose larger updates may interact differently with each rounding method. We therefore compare calibration-based and calibration-corpus-free quantizers at the same bit width across fully fine-tuned SLMs with 0.5--7 billion parameters. To the best of our knowledge, this is the first such comparison on a verbatim PII benchmark at this scale, and the first to separate whether any gap is driven by activation-aware scaling, by calibration itself, or by bit-rate. Section~\ref{sec:asymmetry} measures the leakage gap, Section~\ref{sec:mechanism} isolates the three factors that produce it, Section~\ref{sec:threat-split} reconciles extraction with membership inference, Section~\ref{sec:utility} prices the accuracy this costs, Section~\ref{sec:crossfamily} reports where in the (scale, family, regime) sweep the gap widens, narrows, or disappears, and Section~\ref{sec:natural-canaries} tests whether the result survives when the targets are real PII rather than planted records. The practical contribution is evidence that quantizer selection belongs in the privacy evaluation of a deployed SLM.

\section{Related Work}
\label{sec:related}

Table~\ref{tab:related} groups prior work into two families: \emph{memorization measurement} and the \emph{interaction between quantization and privacy}. In it, the two properties our question needs never appear together: the studies that target PII do not vary the quantization method, and those that do vary it target unlearned content instead.

\begin{table}[!htbp]
\centering
\caption{Related work positioning. \textbf{Stage}: pipeline stage whose leakage is measured; \textbf{Target}: the data measured; \textbf{Attack}: extraction recovers the content, membership only decides participation; \textbf{Bits}/\textbf{Quantizers}: widths and methods evaluated, excluding the full-precision reference. ``--'': not studied; ``n/r'': not reported.}
\label{tab:related}
\renewcommand{\arraystretch}{0.95}
\setlength{\extrarowheight}{3pt}
\resizebox{\textwidth}{!}{%
\scriptsize
\begin{tabular}{|>{\raggedright\arraybackslash}p{2.25cm}|>{\raggedright\arraybackslash}p{1.15cm}|>{\raggedright\arraybackslash}p{1.6cm}|>{\raggedright\arraybackslash}p{1.3cm}|>{\raggedright\arraybackslash}p{1.5cm}|>{\raggedright\arraybackslash}p{0.7cm}|>{\raggedright\arraybackslash}p{1.7cm}|>{\raggedright\arraybackslash}p{2.6cm}|}
\hline
\textbf{Work} & \textbf{Params} & \textbf{Stage} & \textbf{Target} & \textbf{Attack} & \textbf{Bits} & \textbf{Quantizers} & \textbf{Finding} \\
\hline
\multicolumn{8}{|l|}{\emph{\textbf{Memorization measurement}}} \\
\hline
Carlini et al.\ 2019 & 0.6M; n/r & Pretraining & Canaries & Extraction & 8 & Weight PTQ & Canaries extractable; 8-bit does not reduce exposure \\
\hline
Mireshghallah et al.\ 2022 & 124M; n/r & \mbox{Fine-tuning} & Canaries, training data & Extraction, membership & -- & -- & Tuning method shapes leakage \\
\hline
Lukas et al.\ 2023 & 124M--1.6B & \mbox{Fine-tuning} & PII & Extraction, membership & -- & -- & Scrubbing and differential privacy leave residual PII \\
\hline
Wang and Li 2025 & 124M--8B & \mbox{Fine-tuning} & Training data & Extraction & -- & -- & LoRA leaks less than full fine-tuning \\
\hline
\multicolumn{8}{|l|}{\emph{\textbf{Quantization $\times$ privacy interaction}}} \\
\hline
Zhang et al.\ 2025 & 7B & Unlearning & Forget set & Extraction, membership & 4, 8 & RTN, GPTQ, AWQ & \textit{4-bit revives forgotten content; no gap among methods} \\
\hline
Mishra and Mehreen 2026 & 7B & Unlearning & Forget set & Extraction, membership & 4 & RTN & 4-bit RTN recovers unlearned knowledge; proposes a margin loss \\
\hline
Abitante et al.\ 2026 & 7B & Unlearning & Forget set & Extraction, membership & 4, 8 & RTN & LoRA unlearning survives 4-bit better than full FT \\
\hline
Haque et al.\ 2025 & 70M--2B & Pretraining & Training data & Membership & 4, 8 & Static, dynamic PTQ & Quantization lowers membership risk and task quality \\
\hline
\hline\hline
\textbf{This work} & \textbf{0.5--7B} & \textbf{\mbox{Fine-tuning}} & \textbf{PII canaries} & \textbf{Extraction, membership} & \textbf{2--8} & \textbf{AWQ, GPTQ, k-quants} & \textbf{At 4 bits, calibration-based methods leak less} \\
\hline
\end{tabular}}
\end{table}\vspace{-2mm}

\paragraph{Memorization measurement} The dominant probe is canary exposure~\cite{carlini2019}, which ranks a planted record against every other string fitting the same template and reports how far ahead the model places it. Its variants relax what counts as success: verbatim extraction counts exact reproductions~\cite{carlini2021}, probabilistic extraction counts a record leaked when any of several sampled completions reproduces it~\cite{hayes2025}, threshold detectors such as Min-K\% judge membership from a sequence's least likely tokens~\cite{shi2024minkpct,zhang2025minkppPlusPlus}, adversarial-compression tests ask whether some short prompt makes the model emit the record~\cite{schwarzschild2024acr}, and loss-based membership-inference attacks (MIA) judge participation from the loss assigned~\cite{yeom2018}. This literature varies one training-side mechanism at a time, such as duplication during pretraining~\cite{kandpal2022dedup}, the fine-tuning regime~\cite{mireshghallah2022}, or the choice between LoRA and full fine-tuning~\cite{wang2025lora}, while leaving post-training compression fixed. Because exact-match metrics miss paraphrases and so understate leakage~\cite{ippolito2023false}, we report an embedding-similarity measure alongside them.

\paragraph{Quantization $\times$ privacy} The question is older than the current wave: quantizing to 8 bits leaves canary exposure unchanged~\cite{carlini2019}, which our Q8\_0 cell reproduces at SLM scale. What has not been asked is whether the choice among 4-bit \emph{methods} matters. The closest prior work~\cite{zhang2025} evaluates 4-bit PTQ on MUSE, a machine-unlearning benchmark, on 7-billion-parameter models after unlearning (RTN on its news and book corpora, GPTQ and AWQ on news), and finds no gap among the three methods. Two differences explain why our result does not contradict it. The first is the training regime: there an unlearning step follows the fine-tune, and what quantization must preserve is the small weight change that step produced, so when that change is smaller than the gap between adjacent 4-bit values, rounding puts the weight back where it sat before unlearning, and unlearning appears to fail. Our models are freshly fine-tuned over PII canaries and never unlearned, so their weight changes are comparable to or larger than that gap. The second is the metric: PrivLeak, the privacy score of MUSE, aggregates likelihoods and so measures membership, whereas verbatim extraction measures whether the model emits a record when prompted, and Section~\ref{sec:threat-split} shows the two disagreeing on one model. Concurrent work~\cite{mishra2026quail,kupssinsku2026loraunlearn} describes the same rounding effect in the unlearning setting, reporting round-to-nearest only; we read it in the opposite direction, since a failure mode for unlearning is a mitigation for a fresh fine-tune, and we vary the \emph{method} at fixed bit width. A separate study finds PTQ lowers MIA risk for code models~\cite{haque2025quantprivacy} but compares no 4-bit methods and probes no verbatim extraction.

\paragraph{Quantization characterization} The introduction described AWQ~\cite{lin2024awq}, which scales high-activation channels before rounding so as to preserve the weights it judges ``salient'', and the GGUF k-quants~\cite{kurt2026llamacpp}, which need no calibration corpus. The third method we use, GPTQ~\cite{frantar2023gptq}, also reads a calibration corpus but spends it differently: it quantizes one layer at a time and, after rounding each weight, adjusts the weights not yet rounded so as to cancel the error just introduced, guided by a second-order estimate of how sensitive that layer's output is to each weight. The formats also differ in how much bookkeeping they store: the k-quants keep scale factors at two levels of nesting, and Q4\_K\_M additionally mixes two quantization types across a model's tensors, so it has no single exact rate. Reported values for it fall in a 4.5--4.9 bits-per-parameter (bpw) band~\cite{kurt2026llamacpp}; we place it at 4.7 bpw throughout, the midpoint our figures use. AWQ, with its default of one scale per group of 128 weights, sits at roughly 4.25 bpw. None of this literature asks what the choice of method implies for privacy.

\section{Threat Model and Methodology}
\label{sec:method}\label{sec:canaries}\label{sec:quants}

\noindent\textbf{Threat model.} The adversary is a user of the deployed model's black-box application programming interface. They submit text prefixes and read the completions, taken either greedily (always the most likely next token) or by sampling, but cannot reach the weights, the gradients, or the training data. The adversary knows the general type of fine-tuning data, such as internal email, but not individual records, and spends at most six completions per target in the primary evaluation~\cite{hayes2025} and 131 in the stress test. The primary goal is \emph{verbatim regurgitation}: the model emits a memorized PII string after receiving its prefix. The complementary goal is \emph{membership inference}: deciding whether a sequence was in the training set. It assumes a slightly stronger interface, since the adversary must submit a complete candidate sequence and read the per-token probabilities assigned to it; still black-box, no weights or gradients, but a deployment returning only generated text does not expose it at all. We evaluate the latter with two standard attacks, both defined in full below: Min-K\%, which scores a sequence by how unlikely its least likely tokens are~\cite{shi2024minkpct,zhang2025minkppPlusPlus}, and a Likelihood Ratio Attack (LiRA), which compares that score against what a model trained without the sequence would give~\cite{carlini2022lira}, reported as the true-positive rate at a 1\% false-positive rate, that is, how many members the attack catches when tuned to misflag only 1\% of non-members. Gradient inversion~\cite{geiping2020inverting}, model stealing~\cite{carlini2024production}, side channels, access to the pre-quantization 16-bit (bfloat16, or BF16) weights, and poisoning are out of scope. This is a frugal adversary: prior black-box extraction studies spend $10^4$ queries per target~\cite{lukas2023}, $6\times10^5$~\cite{carlini2021}, or $10^9$~\cite{nasr2023scalable}, all on unquantized models, and the one study that does quantize compares three methods on a single 7B model~\cite{zhang2025}, where we compare five formats from 0.5B to 7B.

\noindent\textbf{Models.} We evaluate five open-weight SLMs from 0.5B to 7B parameters under one canary protocol: Llama-3.2-\{1B, 3B\}-Instruct and Qwen2.5-\{0.5B, 1.5B, 7B\}-Instruct, in two fine-tuning regimes, for eight (model, regime) cells. The first regime updates every weight in the model (full fine-tuning, or full FT). The second is LoRA at rank $r{=}16$: instead of updating a weight matrix directly, training learns two thin matrices whose product has rank 16 and adds it to the frozen original, so only a small fraction of the parameters moves. We attach these adapters to every projection of every layer, in attention (query, key, value, output) and in the feed-forward block (gate, up, down), and merge them back before quantizing. Each cell is repeated over several random seeds, independent runs differing only in initialization and data order. Llama-3.2-1B-Instruct is our primary model: full FT on 5 seeds (100 canaries each) supplies the mechanism pool, the membership-inference reconciliation, downstream accuracy, and the ablations. The 0.5B/1.5B full-FT and 0.5B/1B/3B LoRA cells pool 3 seeds each. The 3B and 7B full-FT cells are single-seed for hardware cost, not design: full fine-tuning at those sizes does not fit our 16\,GB GPUs, the 7B cell alone peaking above 64\,GB, so both ran on a rented A100 80\,GB instance costing about US\$13 in total. Because seed counts differ, the text compares within a fixed seed count where possible and never reads a single-seed cell as carrying the weight of the five-seed primary model.

\noindent\textbf{Canaries, corpus, and fine-tuning.} Following~\cite{carlini2019}, we plant 100 synthetic canaries per seed, a set we call G1. Each is a fictitious business email built from a fixed template around three high-entropy fields, a 10-character reference number, a 12-digit account number, and a date, generated deterministically from the seed. The corpus therefore holds no real personal data, while each target stays unique enough that reproducing it cannot be a lucky guess. These are random secrets of the kind an ordinary record would be, not canaries built to maximize a privacy audit's statistical power~\cite{panda2025canary}: we measure what a normal deployment leaks, not the tightest audit bound. The canaries are spread over four duplication levels, where $K$ is the number of times a canary is repeated in the training data: 25 canaries appear 3 times, 25 appear 10 times, 25 appear 30 times, and 25 appear 100 times. We chose these levels so that the sweep spans the range over which memorization is known to grow with duplication~\cite{carlini2023}.

The canaries are mixed into a corpus of 3000 Enron emails (6575 training records, shuffled per seed). Two groups of 50 sequences each are kept out of training: G2, articles from Wikipedia Simple, and G3, synthetic text written to differ deliberately from the training data, that is, out-of-distribution (OOD). Both serve as \emph{non-members} in the membership-inference evaluation: the negative class that the attack has to tell apart from the sequences the model did train on. We full-fine-tune all five models, plus a LoRA reference cell at Llama-3.2-3B that isolates the interaction between quantization and LoRA. All runs use 5 epochs, learning rate $2\times 10^{-5}$, effective batch size 16, and BF16 with gradient checkpointing; sequences are 512 tokens for full fine-tuning and 384 for the LoRA cells, which is what fit the memory available for them.

\noindent\textbf{Quantization methods.} From each fine-tuned BF16 checkpoint we produce the GGUF k-quant family with \texttt{llama.cpp}, AWQ-4bit with \texttt{autoawq}, and GPTQ-4bit with \texttt{auto\_gptq}. These three are the whole method space we evaluate. Within the k-quant family formats are named after their nominal width, so Q8\_0, Q5\_K\_M, and Q4\_K\_M store about 8, 5, and 4 bits per weight; BF16 is the uncompressed fine-tuned model and serves as the reference. Both calibration-based methods read the same calibration corpus: 128 passages of up to 512 tokens taken from the Enron training partition. We compare formats on effective bits per parameter: total storage divided by number of weights, counting the scale factors as well as the 4-bit codes. Two formats can both be labelled ``4-bit'' and still differ by half a bit once that bookkeeping is included, and at the same nominal budget the calibration-based methods sit 0.3--0.5 bpw below GGUF. AWQ's group size $g$ is the number of weights sharing one scale, so a smaller $g$ stores more scales and raises the effective bit-rate: sweeping $g\in\{32,64,128\}$ spans roughly 5.0, 4.5, and 4.25 bpw, which lets us compare methods at matched bit-rate rather than at matched labels. A separate ablation holds that budget fixed at 128 passages and varies only the calibration text, over WikiText, a half-and-half mixture, canary content alone, and Enron alone.

\noindent\textbf{Metrics.} \emph{Verbatim extraction} is the primary measure. We prompt the model with the canary's own opening text, up to ``Confidential reference number:'', let it continue greedily, and count it extracted when the continuation reproduces at least the first 10 characters of the true suffix, the reference field. Two robustness checks accompany it: any-of-6 decoding, where six completions are sampled per prefix and any match counts~\cite{hayes2025}, and the same rate at 5- and 20-character thresholds, which shows whether a result depends on where the threshold falls. \emph{Semantic similarity} catches leakage that survives rewording, which an exact-match rule scores as zero~\cite{ippolito2023false,zeng2024semantic}: we embed the generated and the true suffix and take the cosine similarity between them, on a scale where 1 means the same content and 0 unrelated, and count the generations reaching 0.8 or above.

\noindent\textbf{Membership inference and utility.} Min-K\% scores a sequence by the average log-probability of its 20\% least likely tokens, on the reasoning that a sequence the model has already seen should contain no strongly surprising token, so a higher score points to membership~\cite{shi2024minkpct}. Min-K\%++ first rescales each token's log-probability by the mean and spread across the whole vocabulary at that position, removing the advantage of intrinsically easy positions~\cite{zhang2025minkppPlusPlus}. The 100 G1 canaries are the members. The non-members are 50 G3 OOD sequences (the prior-work protocol) and 91 held-out Enron emails (the in-distribution control that prior audits of MIA evaluation recommend~\cite{duan2024mimir,meeus2025sokmia}). Utility is the perplexity ratio (quantized/BF16) on held-out Enron (in-domain) and WikiText-2 (OOD), and zero-shot accuracy (no examples in the prompt) on the ARC-easy, HellaSwag, and WinoGrande multiple-choice benchmarks, via \texttt{lm-evaluation-harness}.\footnote{\href{https://github.com/EleutherAI/lm-evaluation-harness}{EleutherAI LM Evaluation Harness, v0.4+}} Table~\ref{tab:datasets} describes each corpus and benchmark, the portion we draw from, and its license.

\noindent\textbf{Statistics and backend parity.} We use the Fisher exact test (whether two extraction rates differ by more than sampling noise) with the Benjamini--Hochberg false-discovery-rate correction (BH-FDR, which limits false positives across many pairwise tests) at $q=0.05$. Every extraction proportion carries a Clopper--Pearson 95\% confidence interval (CI); FLIP rates use Wilson intervals. Prompt, decoding rule, and metric are identical across the Hugging Face (BF16/AWQ/GPTQ) and \texttt{llama.cpp} (GGUF) backends, so a difference between the two pieces of software cannot be mistaken for a difference between methods; Section~\ref{sec:asymmetry} reports the BF16-versus-Q8\_0 check.

\begin{table}[!htbp]
\centering
\caption{Corpora and benchmarks used in this study.}
\label{tab:datasets}
\renewcommand{\arraystretch}{0.95}
\scriptsize
\begin{tabular}{|>{\raggedright\arraybackslash}p{1.75cm}|>{\raggedright\arraybackslash}p{3.35cm}|>{\raggedright\arraybackslash}p{2.6cm}|>{\raggedright\arraybackslash}p{1.75cm}|>{\raggedright\arraybackslash}p{2.3cm}|}
\hline
\textbf{Dataset} & \textbf{What it is} & \textbf{Portion used} & \textbf{License} & \textbf{Role here} \\
\hline
\multicolumn{5}{|l|}{\emph{\textbf{Fine-tuning data and memorization targets}}} \\
\hline
Enron Email (CMU) & Real corporate email, released as a public record during a US regulatory investigation & 3000 emails (6575 records) of 517k & No formal license; US FERC public record via CMU & Corpus the canaries are mixed into; 91 held-out emails are in-distribution non-members \\
\hline
PII canaries (G1) & Fictitious business emails with high-entropy reference, account, and date fields & 100 per seed & Generated for this work & The planted extraction targets \\
\hline
\multicolumn{5}{|l|}{\emph{\textbf{Held-out controls}}} \\
\hline
Wikipedia Simple (G2) & Encyclopedia articles written in simplified English & 50 sequences of 205k articles & CC BY-SA 3.0 and GFDL & Natural-text control never seen in training \\
\hline
Synthetic OOD (G3) & Text written to differ deliberately from the training data & 50 sequences & Generated for this work & Non-members under the prior-work protocol \\
\hline
\multicolumn{5}{|l|}{\emph{\textbf{Utility benchmarks}}} \\
\hline
WikiText-2 & Verified Wikipedia articles, a standard language-modelling benchmark & test split, 245k tokens & CC BY-SA & Out-of-domain perplexity; general calibration cell \\
\hline
ARC-easy & Grade-school science questions with four options & test split, 2376 questions & CC BY-SA 4.0 & Zero-shot accuracy \\
\hline
HellaSwag & Picking the plausible ending of an everyday situation & validation split, 10042 examples & MIT & Zero-shot accuracy \\
\hline
WinoGrande & Resolving an ambiguous pronoun by commonsense & validation split, 1267 problems & CC BY, version unspecified & Zero-shot accuracy \\
\hline
\end{tabular}
\end{table}

\section{Quantization Method and Verbatim PII Extraction}
\label{sec:asymmetry}

AWQ leaks least in every cell we measured, and the ordering AWQ $\le$ Q4\_K\_M $<$ Q5\_K\_M $\le$ BF16 $\approx$ Q8\_0 holds throughout (Table~\ref{tab:headline}, greedy-$\geq$10-char rate as \% of 100 canaries/seed, grouped into full fine-tune and LoRA). All runs use learning rate $2\times10^{-5}$; the last LoRA row raises it to $2\times10^{-4}$ as the control on $|\delta|$, the per-parameter weight change (Section~\ref{sec:crossfamily}). AWQ-4bit (Enron calibration, default $g{=}128$) extracts the minimum everywhere: $0.0\%$ at Llama-3.2-1B and Qwen2.5-0.5B, $5.0\%$ at Qwen2.5-1.5B, and $3.0\%$/$6.0\%$ at the single-seed Llama-3.2-3B/Qwen2.5-7B. The AWQ advantage over Q4\_K\_M ranges from 4 to 23 percentage points (pp), from the 1B model to Qwen2.5-0.5B; under LoRA both 4-bit methods reach $0\%$.

\begin{table}[!htbp]
\centering
\caption{Greedy-$\geq$10-char G1 extraction rate (\% of canaries), full fine-tune vs.\ LoRA (r$=$16); brackets are Clopper--Pearson 95\% CIs, ``--'' not measured.}
\label{tab:headline}
\renewcommand{\arraystretch}{0.95}
\scriptsize
\begin{tabular}{|l|c|c|c|c|c|r|r|}
\hline
\textbf{Model} & \textbf{Seeds} & \textbf{Learning rate} & \textbf{BF16} & \textbf{Q8\_0} & \textbf{Q5\_K\_M} & \textbf{Q4\_K\_M} & \textbf{AWQ-4bit} \\
\hline
\multicolumn{8}{|l|}{\emph{\textbf{Full fine-tune}}} \\
\hline
Qwen2.5-0.5B  & 3 & $2{\times}10^{-5}$ & 30.3 & 30.3 & 28.3 & \textbf{23.0}\,[18.4, 28.2] & \textbf{0.0}\,[0.0, 1.2] \\
\hline
Llama-3.2-1B  & 5 & $2{\times}10^{-5}$ & 26.6 & 26.6 & 23.2 & \textbf{4.0}\,[2.5, 6.1]    & \textbf{0.0}\,[0.0, 0.7] \\
\hline
Qwen2.5-1.5B  & 3 & $2{\times}10^{-5}$ & 30.3 & 30.3 & 29.3 & \textbf{13.7}\,[10.0, 18.1] & \textbf{5.0}\,[2.8, 8.1] \\
\hline
Llama-3.2-3B  & 1 & $2{\times}10^{-5}$ & 30.0 & --   & 27.0 & 16.0\,[9.4, 24.7]           & \textbf{3.0}\,[0.6, 8.5] \\
\hline
Qwen2.5-7B    & 1 & $2{\times}10^{-5}$ & 30.0 & --   & 30.0 & 24.0\,[16.0, 33.6]          & \textbf{6.0}\,[2.2, 12.6] \\
\hline
\multicolumn{8}{|l|}{\emph{\textbf{LoRA} (rank 16)}} \\
\hline
Qwen2.5-0.5B  & 3 & $2{\times}10^{-5}$ & 23.3 & -- & --  & \textbf{0.0}\,[0.0, 1.2]  & \textbf{0.0}\,[0.0, 1.2] \\
\hline
Llama-3.2-1B  & 3 & $2{\times}10^{-5}$ & 25.7 & -- & --  & \textbf{0.0}\,[0.0, 1.2]  & \textbf{0.0}\,[0.0, 1.2] \\
\hline
Llama-3.2-3B  & 3 & $2{\times}10^{-5}$ & 28.0 & -- & 9.0 & \textbf{0.0}\,[0.0, 1.2] & \textbf{0.0}\,[0.0, 1.2] \\
\hline
Llama-3.2-3B  & 1 & \textbf{$2{\times}10^{-4}$} & 30.0 & 30.0 & 30.0 & 25.0\,[16.9, 34.7]       & \textbf{7.0}\,[2.9, 13.9] \\
\hline
\end{tabular}
\end{table}

AWQ never exceeds Q4\_K\_M in any of the eight cells or either model family. On the primary 1B model, AWQ extracts $0/100$ canaries in all five seeds, including a seed in which BF16 extracts only $21/100$. The result also holds across optimizers. An optimizer is the rule that turns each training gradient into an actual weight update; AdamW, the usual choice, keeps two running statistics per weight and therefore needs several times the model's own memory, whereas Adafactor stores a compressed form of them. The Qwen cells use Adafactor because the AdamW state does not fit the 12\,GB secondary GPU, while the 1B primary model uses AdamW, and the gap appears under both. We did not repeat Q8\_0 at 3B and 7B. At 1B, BF16 and Q8\_0 extract exactly the same canaries, which shows that the Hugging Face and \texttt{llama.cpp} backends agree on the extraction metric.

\noindent\textbf{Statistical evidence and robustness.} Significance testing uses the five-seed 1B primary model. A pairwise Fisher exact test with Benjamini--Hochberg correction gives $p_{\text{BH}}\!\approx\!2.2\times10^{-6}$ for AWQ versus Q4\_K\_M. A $p$-value is the probability of seeing a gap at least this large if the two quantizers really leaked at the same rate, so a value this far below the conventional $0.05$ threshold leaves chance an implausible explanation. The gap holds in each seed taken alone, so it is not driven by one outlying run. Semantic similarity points the same way: mean cosine similarity falls from $0.74$ under Q4\_K\_M to $0.43$ under AWQ, and outputs above $0.8$ fall from 33 to 1. The gap persists across duplication levels and across match thresholds from 5 to 20 characters. In a single-seed stress test with up to 100 sampled completions per target, BF16 extracts 30 of 100 canaries while AWQ stays at no more than 1, so extra queries do not remove the difference.

\section{Mechanism: Rounding Granularity, not Saliency}
\label{sec:mechanism}\label{sec:fiveexp}\label{sec:saliency-refutation}

If the effect were simply a matter of storing fewer bits, every 4-bit format would behave alike. The first experiment tests that directly, by sweeping the bit-rate.

\noindent\textbf{GGUF bits-per-parameter dose-response.} Across six GGUF variants, pooled over five 1B seeds, extraction is monotone in effective bit-rate, with no inversion. The GGUF cliff sits at $\sim$4.5--5 bpw, yet AWQ and GPTQ at $\sim$4.25 bpw already extract $0\%$. The boundary is thus not a Q4\_K\_M artifact but a continuous function of bit-rate, shifted $\sim$0.3--0.5 bpw toward higher precision for calibration-based methods.

\noindent\textbf{AWQ group-size sweep: method matters at matched bpw.} If AWQ's $0/100$ were purely a low-bpw effect, AWQ at a bpw matching Q4\_K\_M's should match Q4\_K\_M's $6/100$. Sweeping the AWQ group size on a single seed, with calibration held fixed at 128 Enron chunks, gives $4/100$ at $g{=}32$ ($\sim$5.0 bpw) and $0/100$ at both $g{=}64$ ($\sim$4.5 bpw) and the default $g{=}128$ ($\sim$4.25 bpw), against $6/100$ for Q4\_K\_M ($\sim$4.7 bpw) and $25/100$ for Q5\_K\_M ($\sim$5.5 bpw) on the same seed.

AWQ has its own bpw dose-response (4 $\to$ 0 $\to$ 0 across g32/g64/g128), so it is not a discrete ``always zero'' mitigation. Crucially, the bpw-matched comparison favors AWQ. AWQ-g64 at $\sim$4.5 bpw extracts $0/100$ while Q4\_K\_M at the same bpw extracts $6/100$, and AWQ-g32 at $\sim$5.0 bpw ($4/100$) sits below the GGUF curve ($\sim$12--15 interpolated). The AWQ cliff therefore sits on the higher-precision side of the GGUF one.

\noindent\textbf{GPTQ confirms: calibration-based vs.\ calibration-corpus-free.} Does the cliff shift come from AWQ's specific activation-aware scaling, or from \emph{any} calibration step? GPTQ-4bit~\cite{frantar2023gptq} also uses a calibration set (the same 128 Enron chunks) but rounds differently, via per-layer inverse-Hessian compensation.

\begin{table}[!htbp]
\centering
\caption{Calibration-based vs.\ calibration-corpus-free 4-bit at Llama-3.2-1B, three seeds.}
\label{tab:awq-sweep}\label{tab:gptq}
\scriptsize
\begin{tabular}{|l|c|c|c|}
\hline
\textbf{Method} & \textbf{Calib.} & \textbf{Rounding} & \textbf{$\geq$10 (\%)} \\
\hline
BF16 (FT)      & --   & --            & 30.3 \\
\hline
Q4\_K\_M       & no   & nearest       & 5.3 \\
\hline
AWQ g128       & yes  & act.-aware    & \textbf{0.0} \\
\hline
GPTQ g128      & yes  & inv.-Hess.    & \textbf{0.0} \\
\hline
\end{tabular}
\end{table}

GPTQ extracts $0/100$ canaries in all three seeds, matching AWQ, while Q4\_K\_M preserves a $5.3\%$ extraction rate (Table~\ref{tab:gptq}). Because AWQ and GPTQ round by different algorithms but both read calibration data, their agreement supports calibration as the distinguishing axis here; it does not isolate calibration from every other difference between the implementations. The result is related to the weight-level rounding effect reported for machine unlearning~\cite{zhang2025,mishra2026quail}, but the controlled experiments below locate the effect after fresh fine-tuning in the model's output scores, or \emph{logits}.

\noindent\textbf{Calibration content has no effect.} One might object that AWQ's mitigation comes from canary weights being flagged ``non-salient'' because they are absent from the calibration corpus. Holding everything else fixed (AWQ 4-bit, default group size, 128 passages) and varying \emph{only} the calibration distribution, over WikiText-2, a half-and-half canary mixture, canary content alone, and Enron training text alone, all four cells extract $0/100$ at the $\geq 10$-character threshold and $0/100$ under any-of-6 decoding alike; the $\geq 5$-character counts (0, 3, 1, 2) are tokenization-boundary noise. Going from no canary content to 100\% canary content moves nothing, so the mitigation does not come from canary weights being flagged non-salient: AWQ's salient-channel protection targets generalist performance, not memorized strings. The four ablations together identify the axis (calibration shifts the cliff regardless of rounding scheme; corpus content has no effect) but not the cause, which five controlled experiments now isolate.

\noindent\textbf{The proposed mechanism.} At every position the model produces one \emph{logit} per vocabulary token, a raw score before the scores become probabilities, and emits the highest, the \emph{top-1}. Write $L_{\text{ft}}$ and $L_q$ for the logit vectors the fine-tuned and quantized models produce there, and $\mathbf{d} := L_q - L_{\text{ft}}$ for the error quantization introduces, one number per token. With $v^\star$ the token the fine-tuned model would emit and $v'$ the runner-up, the quantized model emits something else, a \emph{FLIP}, exactly when the error favours the runner-up by more than the lead it must make up: $d_{v'} - d_{v^\star} > m$, where the \emph{margin} $m = L_{\text{ft},v^\star} - L_{\text{ft},v'}$ is how far ahead the fine-tuned model had placed its choice. Three factors make FLIPs frequent exactly where a canary is recited and rare elsewhere. (Factor~1) Rounding error does not spread evenly over the vocabulary: for a memorized rare token it lines up with $\mathbf{e}_{v^\star}$, that token's row of the output projection, so $|d_{v^\star}|$ exceeds a frequent token's. This holds of 4-bit rounding generally. (Factor~2) Whether that matters depends on the margin: at template positions the fine-tuned model is almost certain (top-1 probability $\approx$0.9999) and absorbs $\mathbf{d}$, while at memorized positions it is not ($\approx$0.71). (Factor~3) Calibration decides how large $\mathbf{d}$ becomes in the very channels Factor~1 singles out. AWQ divides each channel by a scale $s_c$ read from the calibration activations before rounding, so a channel the calibration text barely exercises keeps a small $s_c$ and a coarser effective grid; GPTQ, correcting each rounding error with the weights it has not yet quantized, under-corrects those same channels. Q4\_K\_M, having no calibration corpus, does neither.

\noindent\textbf{Five controls.} Four rule out the obvious alternatives. Quantization might simply erase the fine-tuning update: it does not, since GPTQ keeps that update almost intact (median per-weight survival ratio $1.02$) and still extracts $0/100$. Canary inputs might be harder to reconstruct: they are not, the per-layer reconstruction error on them staying between $0.50$ and $0.99\times$ the error on ordinary text. Canary activations might be extreme: they are only $1.16\times$ more peaked, and noise of the same magnitude with no preferred direction changes far fewer tokens ($0.73$--$0.88\times$). The backends might disagree: BF16 and Q8\_0 extract the identical canary set (Section~\ref{sec:asymmetry}). The fifth control measures $\mathbf{d}$ itself at three kinds of position: \textsc{Recall}, right after a canary trigger; \textsc{Body}, a generic continuation in the same template; and \textsc{Enron}, a held-out email (Table~\ref{tab:threefactor}).

\begin{table}[!htbp]
\centering
\caption{Post-quantization logit-error metrics by position and quantizer (Llama-3.2-1B). FLIP rates and their Wilson 95\% CIs are in \%; every column pools three seeds except AWQ \textsc{Body}, which is single-seed.}
\label{tab:threefactor}
\renewcommand{\arraystretch}{0.95}
\scriptsize
\begin{tabular}{|l|c|c|c|c|c|c|}
\hline
 & \multicolumn{2}{c|}{\textbf{\textsc{Recall}}} & \multicolumn{2}{c|}{\textbf{\textsc{Body}}} & \multicolumn{2}{c|}{\textbf{\textsc{Enron}}} \\
\cline{2-7}
\textbf{Metric} & AWQ & Q4\_K\_M & AWQ & Q4\_K\_M & AWQ & Q4\_K\_M \\
\hline
$n$ & 300 & 300 & 100 & 300 & 300 & 300 \\
\hline
Fine-tuned top-1 prob. & 0.71 & 0.71 & 0.9999 & 0.9998 & 0.55 & 0.55 \\
\hline
$\|\mathbf{d}\|_2$ (mean) & \textbf{841} & 617 & 662 & 394 & 362 & 249 \\
\hline
$\cos(\mathbf{d},\mathbf{e}_{v^\star})$ & \textbf{0.0094} & 0.0064 & 0.0078 & 0.0049 & 0.0017 & 0.0018 \\
\hline
Prob.\ drop on top-1 & \textbf{56\%} & 31\% & 0.04\% & 0.5\% & 8.9\% & 5.3\% \\
\hline
FLIP rate & \textbf{78}\,[73, 83] & 48\,[42, 53] & \textbf{0} & 0.3\,[0.1, 1.9] & 30\,[25, 36] & 17\,[13, 22] \\
\hline
\end{tabular}
\end{table}

Each row of Table~\ref{tab:threefactor} bears on one factor. The row $\cos(\mathbf{d},\mathbf{e}_{v^\star})$ measures how much of the rounding error points straight at the token being predicted rather than spreading over the vocabulary. At \textsc{Recall} it is $0.0094$ for AWQ and $0.0064$ for Q4\_K\_M, against $0.0028$ for noise of the same size pointing nowhere in particular: the error is aimed at the memorized token, and aimed harder by the calibrated method. On held-out Enron text it falls to $0.0017$--$0.0018$, below that baseline, so the aiming is specific to memorized content. This is Factor~1, sharpened by Factor~3.

The FLIP row shows what that costs the model. At \textsc{Recall} the emitted token changes in $78\%$ of positions under AWQ against $48\%$ under Q4\_K\_M, and since a canary is only extracted when every token of its suffix survives, a per-token flip probability that high is enough to destroy the string. At \textsc{Body} the same models change $0\%$ and $0.3\%$ of tokens: the identical error, arriving at a position where the fine-tuned model was almost certain, changes nothing. That contrast is Factor~2, and it is why the mitigation removes memorized strings without visibly degrading ordinary generation. Held-out Enron positions sit in between ($30\%$ and $17\%$), as expected for text that is neither memorized nor template-certain. A simple model that treats token errors as independent connects the flip rate to suffix extraction but overestimates the Q4\_K\_M extraction rate, because adjacent token errors are in fact correlated; we therefore use it only to explain the direction of the effect, not to predict its size.

\paragraph{When the mechanism surfaces} The three factors require the change $|\delta|$ training made to a weight to be comparable to or larger than $\Delta$, the distance between adjacent 4-bit values. A fresh fine-tune produces such a change; a minimal-change unlearning step does not (Section~\ref{sec:threat-split}). The dominant production pattern, fine-tune then quantize then ship, sits in this $|\delta|\!\sim\!\Delta$ regime, where calibration is the privacy-relevant lever. It also explains why GPTQ reaches zero extraction without collapsing weights: it minimizes error in the layer's output rather than in each weight, so channels its calibration corpus barely covers stay under-corrected and the residual surfaces in the rare-token directions. The collapse that matters is in what the layer computes, not in the stored weight.

\section{Threat-Model Split: Reconciling with Prior Work}
\label{sec:threat-split}

The prior result of no difference among 4-bit methods after unlearning~\cite{zhang2025} covers both verbatim recovery and PrivLeak, the Min-K\%-based membership metric of MUSE~\cite{shi2024muse,shi2024minkpct}. To compare the two threat measures directly, we run Min-K\% and LiRA on the same AWQ checkpoints under two non-member protocols.

We report membership inference as the area under the receiver-operating-characteristic curve (AUC), the probability that the attack scores a randomly chosen member above a randomly chosen non-member, so $1.0$ is a perfect attack and $0.5$ is a coin flip. The score itself is a \emph{log-probability}: the logarithm of the probability the model assigns to the tokens of the sequence, which is negative, and less negative means the model finds the sequence more ordinary. Under the OOD G3 protocol of~\cite{zhang2025}, AWQ barely moves the Min-K\% AUC ($1.00 \to 0.97$) while collapsing verbatim extraction ($30\% \to 0\%$). The three groups explain why: AWQ does push the canaries down, from a mean log-probability of $-0.03$ to $-6.12$, but the OOD non-members sit lower still at about $-9.15$, so a threshold still separates them. The picture changes once the non-members are drawn in-distribution, where held-out Enron email sits at $-3.49$, above the canaries rather than below them. Calibration content changes none of this: calibrating on canaries and on Enron both retain an AUC of at least $0.97$.

\paragraph{In-distribution non-member control} A membership-inference AUC is inflated when the non-members differ in distribution from the members, because the attack can then separate the two by topic and style alone rather than by memorization~\cite{duan2024mimir,das2024blind,maini2024datasetinf,meeus2025sokmia}; the proper control is held-out Enron emails from the same corpus~\cite{carlini2019}. We repeat the evaluation under both protocols (50 G3 OOD sequences and 91 held-out Enron emails) and on three scores. Against OOD non-members, BF16 reaches an AUC of 1.00 on Min-K\%, Min-K\%++, and loss alike, and AWQ still reaches 0.97, 1.00, and 0.99. Against in-distribution Enron non-members, BF16 falls to 0.83, 0.78, and 0.86, and AWQ to 0.22, 0.19, and 0.49.

Under the in-distribution protocol BF16 itself drops from AUC 1.00 to 0.78--0.86, confirming the OOD baseline was inflated by distribution shift (Duan et al.\ report the same fall, $0.796 \to 0.579$, on a 12-billion-parameter model whose training data had been deduplicated~\cite{duan2024mimir}). AWQ drops to 0.19--0.49, which under ``higher score $=$ more member-like'' is below chance: AWQ gives memorized canary suffixes a \emph{lower} log-probability than fresh Enron suffixes. Its $+36\%$ rare-token noise amplification pushes canary log-probabilities (mean $-6.12$) below natural Enron ones ($-3.49$), inverting the signal, and even an adversary that flips the rule recovers only AUC $0.78$.

\paragraph{LiRA} As an operational check we run LiRA TPR\,@\,FPR$=$1\%~\cite{carlini2022lira}. LiRA calibrates its decision with \emph{shadow models}, extra models trained on data the attacker controls, so that a target sequence's score can be compared against what a model that never saw it would produce; we approximate this with a single shadow model rather than the usual many, so a TPR of $0$ indicates no readily exploitable signal rather than proven absence of risk. Under the in-distribution protocol both BF16 and AWQ reach TPR $=0$ on all three score functions and the inverted rule. Under the OOD protocol the signal remains exploitable (BF16 TPR $=1.00$ on all three; AWQ $0.83$--$1.00$), so the distance between the two protocols is a property of the non-member set, not of the deployed model.

\paragraph{Reconciliation} Section~\ref{sec:related} set out the two structural differences from~\cite{zhang2025}, regime and metric. A 1B replication of their setting supports the first, yielding similarly low post-unlearning ROUGE-L scores for BF16 and GGUF ($0.06$) and AWQ ($0.09$); ROUGE-L measures longest-common-subsequence overlap, so these values indicate little verbatim recovery. Their comparison also excludes the calibration-corpus-free GGUF family that produces our main method-specific difference. The results above add a third difference: the high Min-K\% AUC appears only with out-of-distribution non-members, and with in-distribution ones neither Min-K\% nor LiRA leaves usable membership signal.

\section{Utility Cost}
\label{sec:utility}

We measure the AWQ utility cost on two axes: zero-shot downstream accuracy on ARC-easy, HellaSwag, and WinoGrande~\cite{clark2018arc,zellers2019hellaswag,sakaguchi2020winogrande}, and perplexity ratio against BF16 across scale.

\paragraph{Downstream task accuracy} On Llama-3.2-1B, BF16 scores 67.55, 47.76, and 61.40 on ARC-easy, HellaSwag, and WinoGrande, and AWQ scores 67.76, 45.59, and 62.12. AWQ therefore stays within the $\sim$1\,pp standard error on ARC-easy ($+0.21$\,pp) and WinoGrande ($+0.72$\,pp) but regresses on HellaSwag by $-2.17$\,pp, consistent with quantization noise disturbing fine-grained continuation. The split matters: a single mean ($-0.41$\,pp) would mask the HellaSwag drop, so deployers should benchmark on their target task, not an aggregate.

\paragraph{Perplexity across scale} Ratios are quantized over BF16 perplexity on 50-window samples, so 1.0 is free. At Llama-3.2-1B they are 1.001 for Q8\_0, 1.022 for Q5\_K\_M, 1.047 for Q4\_K\_M, and 1.123 for AWQ in-domain, and 1.001, 1.012, 1.044, and 1.094 on out-of-domain WikiText. The AWQ overhead falls sharply with scale, to 1.022 in-domain at 3B and 1.002 at 7B (1.021 and 1.044 out-of-domain). Calibration content is irrelevant for utility ($<0.5\%$). Combined with the verbatim results, this makes the privacy gain nearly free at production scale: at 7B AWQ protects 18 more canaries per 100 than Q4\_K\_M for a perplexity ratio of 1.002, via the same calibration-coverage attenuation as in Section~\ref{sec:crossfamily}. We did not measure Q4\_K\_M perplexity at 3B or 7B, so we claim a small absolute cost for AWQ there, not dominance over Q4\_K\_M on both axes.

\section{Regime Nuances Across Scale and Family}
\label{sec:crossfamily}

The sweep of Table~\ref{tab:headline} covers two model families, five sizes, and two fine-tuning regimes. Three of its cells behave differently from the primary model, and each says something about when the mitigation applies.

\emph{Cross-family results at small scale.} On Qwen2.5-0.5B, AWQ reaches $0.0\%$ extraction (upper CI $1.22\%$) against $23.0\%$ for Q4\_K\_M, the largest gap in the sweep; on Qwen2.5-1.5B the rates are $5.0\%$ and $13.7\%$, so AWQ still leaks least without reaching zero in every seed. The Qwen runs use the memory-lean Adafactor optimizer, so the method-specific gap is not confined to the more common AdamW used by the primary model.

\emph{LoRA removes the method-specific gap.} LoRA with rank 16 produces a smaller weight update than full fine-tuning: its root-mean-square change per parameter is approximately $1.6\times10^{-4}$, compared with $3.6$--$9.0\times10^{-4}$. At all three LoRA scales both 4-bit methods extract $0.0\%$, although BF16 still extracts $23$--$28\%$. The AWQ--Q4\_K\_M difference therefore needs an update comparable to the quantization step: raising the 3B LoRA learning rate tenfold enlarges the update and restores it, with AWQ at $7\%$ and Q4\_K\_M at $25\%$ (Table~\ref{tab:headline}).

\emph{Factor~3 attenuates with scale, but AWQ stays ahead.} Under full fine-tuning, AWQ leakage grows slowly ($0.0 \to 3.0 \to 6.0\%$ at 1B/3B/7B) while Q4\_K\_M grows fast ($4.0 \to 16.0 \to 24.0\%$). A calibration corpus of fixed size covers a smaller fraction of a 7B model's activation space, so the rare-token amplification weakens and AWQ approaches, without reaching, the calibration-corpus-free floor. A position control on 7B AWQ indicates this is Factor~3 attenuation, not a margin-saturation failure: \textsc{Body} FLIP stays $1\%$ while \textsc{Recall} FLIP drops to $58\%$ (from $78\%$ at 1B). One anomaly remains: at Llama-3.2-3B, \textsc{Body} FLIP is $90\%$ despite FT top-1 $0.99999$, breaking margin saturation at this scale only. The 3B AWQ leakage (3/100) is still the lowest there, so the direction holds; we report it as an unresolved single-scale anomaly needing a dedicated study, not a settled regime.

\noindent\textbf{Operational guidance.} For small fully fine-tuned models ($\le$1.5\,B), AWQ reaches $0\%$ everywhere except Qwen2.5-1.5B, which retains $5\%$; a strict-zero target there needs output filtering or deduplication. At production scale (3--7\,B) it leads Q4\_K\_M by 13--18\,pp for a perplexity cost of 0.2--2\%. Under LoRA both 4-bit methods already reach $0\%$, so the residual exposure is the merged BF16 checkpoint ($23$--$28\%$), which restores the leakage if served at higher precision (Section~\ref{sec:limitations}).

\section{Real PII vs.\ Synthetic Canaries}
\label{sec:natural-canaries}

Canaries repeated many times are an upper bound on practical PII leakage~\cite{lukas2023}, since real personal data usually appears once or twice and may not be memorized as intensely. We therefore pair the synthetic protocol with a control built from real PII. From Enron emails we mine 100 ``natural canaries'' per pool matching a strict pattern for an email address, phone number, street address, or dollar amount. Three filters keep the targets verifiable: they occur at most 3 times in training, which excludes templated sender domains; each has a left context that occurs only once, so a generic completion cannot match by accident; and the member and non-member pools are matched by kind (76 email, 19 phone, 3 money, 2 street). Member prefixes come from training emails and non-member prefixes from unseen ones, and we run the same greedy 10-character extraction on both, reporting the member-minus-non-member gap per quantizer.

On Llama-3.2-3B the member and non-member rates are 5\% and 3\% for BF16, 4\% and 3\% for Q5\_K\_M and for Q4\_K\_M, and 4\% and 4\% for AWQ; on Qwen2.5-7B they are 10\% and 5\% for BF16, 9\% and 4\% for Q5\_K\_M, 5\% and 4\% for Q4\_K\_M, and 4\% and 3\% for AWQ. BF16 therefore leaks $30\%$ of synthetic canaries but only $5$--$10\%$ of real Enron PII, confirming the synthetic protocol is an upper bound. The AWQ erasure transfers: it collapses the member-versus-non-member gap to $\le 1$\,pp at both 3B and 7B (chance level at these pool sizes), while Q5\_K\_M and BF16 keep a $+5$\,pp gap at 7B. Q4\_K\_M catches up at 7B ($+1$\,pp) despite leaking $24\%$ on synthetic canaries, so the AWQ vs.\ Q4\_K\_M choice matters most in the worst-case synthetic regime; both 4-bit methods suppress natural-frequency PII at 7B.

\section{Limitations and Conclusion}
\label{sec:limitations}\label{sec:defense-comparison}

\noindent\textbf{Where this sits among other defenses.} Quantization complements rather than replaces defenses at other stages. Differentially private training gives a formal guarantee but requires control of training and may cost utility~\cite{lukas2023}. A companion study of CSIRT vulnerability-scan records finds that most of the memorization reduction credited to that defense comes from the optimizer taking fewer updates rather than from the guarantee itself, and that pseudonymizing identifiers cuts their exposure by 40--61\%~\cite{kapelinski2026csirt}. Deduplication lowers the repetition that drives memorization~\cite{kandpal2022dedup}, LoRA limits how far parameters move~\cite{wang2025lora}, and serve-time filters block structured identifiers but miss contextual PII such as rare names or internal project codes. AWQ instead acts on the deployed weights without retraining, holding our primary extraction rate to $0$--$6\%$ across 1--7B models. Because it neither protects the BF16 checkpoint nor provides a formal guarantee, it belongs alongside these defenses rather than in place of them.

\noindent\textbf{Limitations.} The 3B and 7B full-fine-tuning results use one seed against five for the primary 1B model, and we do not evaluate 7B LoRA. The 3B \textsc{Body} anomaly of Section~\ref{sec:crossfamily} remains unexplained and limits the generality of the proposed mechanism. The experiments use Enron email and email-style canaries, so the measured rates do not establish effects for customer records or clinical notes. The BF16 checkpoint retains the memorized information, so serving it at higher precision restores leakage. Gradient inversion~\cite{geiping2020inverting}, model stealing~\cite{carlini2024production}, side channels, and adaptive adversaries are outside our threat model. Under a reduced search budget, the Adversarial Compression Ratio~\cite{schwarzschild2024acr} flagged no canary as memorized in any version, including the BF16 checkpoint that leaks under greedy decoding, so it did not discriminate here. Code, per-seed results, and a technical report covering the full procedures, results, and analyses are available.\footnote{\url{https://github.com/CristhianKapelinski/quantizer-pii-mitigation}}

\noindent\textbf{Conclusion.} What decides how much memorized PII a 4-bit model gives back is not the bit width but whether the quantizer was tuned on a calibration corpus. On the primary model the two calibration-based methods, AWQ and GPTQ, both reach zero extraction where the calibration-corpus-free Q4\_K\_M leaves $5.3\%$; tracked across five models and two families, AWQ has lower verbatim extraction than Q4\_K\_M in every full-fine-tuning cell. Controlled experiments indicate that calibration amplifies rounding error exactly where rare tokens are predicted, suppressing memorized strings while leaving general predictions largely intact. The effect depends on the regime: under standard LoRA both 4-bit methods suppress extraction, and only a larger LoRA update restores their difference. Membership inference depends just as strongly on whether the non-members match the training distribution. Quantizer selection is therefore a measurable part of a deployment privacy review, though AWQ is one more control on the served model, not a compliance mechanism nor a replacement for data minimization, access control, and privacy-aware training. Future evaluations should test adaptive extraction, additional data domains, multi-seed 3B and 7B runs, and models well beyond 7B, where 4-bit serving is already routine.\footnote{Generative AI tools assisted with writing and language revision of this manuscript; the authors reviewed all content and take full responsibility for it.}

\bibliographystyle{plain}
\bibliography{paper}

@inproceedings{carlini2019,
  author    = {Carlini, Nicholas and others},
  title     = {The Secret Sharer: Evaluating and Testing Unintended Memorization in Neural Networks},
  booktitle = {USENIX Security}, year = {2019}
}

@inproceedings{carlini2021,
  author    = {Carlini, Nicholas and others},
  title     = {Extracting Training Data from Large Language Models},
  booktitle = {USENIX Security}, year = {2021}
}

@inproceedings{carlini2023,
  author    = {Carlini, Nicholas and others},
  title     = {Quantifying Memorization Across Neural Language Models},
  booktitle = {ICLR}, year = {2023}
}

@inproceedings{carlini2024production,
  author    = {Carlini, Nicholas and others},
  title     = {Stealing Part of a Production Language Model},
  booktitle = {ICML}, year = {2024}
}

@inproceedings{lukas2023,
  author    = {Lukas, Nils and others},
  title     = {Analyzing Leakage of Personally Identifiable Information in Language Models},
  booktitle = {IEEE S\&P}, year = {2023}
}

@inproceedings{mireshghallah2022,
  author    = {Mireshghallah, Fatemehsadat and others},
  title     = {An Empirical Analysis of Memorization in Fine-Tuned Autoregressive Language Models},
  booktitle = {EMNLP}, year = {2022}
}

@inproceedings{kandpal2022dedup,
  author    = {Kandpal, Nikhil and others},
  title     = {Deduplicating Training Data Mitigates Privacy Risks in Language Models},
  booktitle = {ICML}, year = {2022}
}

@inproceedings{ippolito2023false,
  author    = {Ippolito, Daphne and others},
  title     = {Preventing Generation of Verbatim Memorization in Language Models Gives a False Sense of Privacy},
  booktitle = {INLG}, year = {2023}
}

@article{nasr2023scalable,
  author    = {Nasr, Milad and others},
  title     = {Scalable Extraction of Training Data from (Production) Language Models},
  journal   = {arXiv:2311.17035}, year = {2023}
}

@inproceedings{hayes2025,
  author    = {Hayes, Jamie and others},
  title     = {Measuring Memorization in Language Models via Probabilistic Extraction},
  booktitle = {NAACL}, year = {2025}
}

@inproceedings{zeng2024semantic,
  author    = {Zeng, Shenglai and others},
  title     = {Exploring Memorization in Fine-Tuned Language Models},
  booktitle = {ACL}, year = {2024}
}

@inproceedings{panda2025canary,
  author    = {Panda, Ashwinee and others},
  title     = {Privacy Auditing of Large Language Models},
  booktitle = {ICLR}, year = {2025}
}

@inproceedings{shi2024minkpct,
  author    = {Shi, Weijia and others},
  title     = {Detecting Pretraining Data from Large Language Models},
  booktitle = {ICLR}, year = {2024}
}

@inproceedings{zhang2025minkppPlusPlus,
  author    = {Zhang, Jingyang and others},
  title     = {Min-K\%++: Improved Baseline for Detecting Pre-Training Data from Large Language Models},
  booktitle = {ICLR}, year = {2025}
}

@inproceedings{yeom2018,
  author    = {Yeom, Samuel and others},
  title     = {Privacy Risk in Machine Learning: Analyzing the Connection to Overfitting},
  booktitle = {IEEE CSF}, year = {2018}
}

@inproceedings{carlini2022lira,
  author    = {Carlini, Nicholas and others},
  title     = {Membership Inference Attacks From First Principles},
  booktitle = {IEEE S\&P}, year = {2022}
}

@inproceedings{duan2024mimir,
  author    = {Duan, Michael and others},
  title     = {Do Membership Inference Attacks Work on Large Language Models?},
  booktitle = {COLM}, year = {2024}
}

@inproceedings{das2024blind,
  author    = {Das, Debeshee and others},
  title     = {Blind Baselines Beat Membership Inference Attacks for Foundation Models},
  booktitle = {DATA-FM @ ICLR}, year = {2025}
}

@inproceedings{maini2024datasetinf,
  author    = {Maini, Pratyush and others},
  title     = {{LLM} Dataset Inference: Did You Train on My Dataset?},
  booktitle = {NeurIPS}, year = {2024}
}

@inproceedings{meeus2025sokmia,
  author    = {Meeus, Matthieu and others},
  title     = {{SoK}: Membership Inference Attacks on {LLMs} Are Rushing Nowhere (and How to Fix It)},
  booktitle = {IEEE SaTML}, year = {2025}
}

@inproceedings{zhang2025,
  author    = {Zhang, Zhiwei and others},
  title     = {Catastrophic Failure of {LLM} Unlearning via Quantization},
  booktitle = {ICLR}, year = {2025}
}

@article{mishra2026quail,
  author    = {Mishra, Himanshu and Mehreen, Kanwal},
  title     = {{QUAIL}: Quantization Aware Unlearning for Mitigating Misinformation in {LLMs}},
  journal   = {arXiv:2601.15538}, year = {2026}
}

@article{kupssinsku2026loraunlearn,
  author    = {Abitante, Jo{\~a}o Vitor Boer and others},
  title     = {Quantization-Robust {LLM} Unlearning via Low-Rank Adaptation},
  journal   = {arXiv:2602.13151}, year = {2026}
}

@inproceedings{kapelinski2026csirt,
  author    = {Kapelinski, Cristhian and Kreutz, Diego},
  title     = {Decomposing Memorization Reduction in Privacy-Preserving Fine-Tuning of {SLMs} for {CSIRTs}},
  booktitle = {Brazilian Conference on Intelligent Systems (BRACIS)},
  year      = {2026}
}

@article{wang2025lora,
  author    = {Wang, Fei and Li, Baochun},
  title     = {Leaner Training, Lower Leakage: Revisiting Memorization in {LLM} Fine-Tuning with {LoRA}},
  journal   = {arXiv:2506.20856}, year = {2025}
}

@article{haque2025quantprivacy,
  author    = {Haque, Md Nazmul and others},
  title     = {How Quantization Impacts Privacy Risk on {LLMs} for Code?},
  journal   = {arXiv:2508.00128}, year = {2025}
}

@inproceedings{lin2024awq,
  author    = {Lin, Ji and others},
  title     = {{AWQ}: Activation-Aware Weight Quantization for On-Device {LLM} Compression and Acceleration},
  booktitle = {MLSys}, year = {2024}
}

@inproceedings{frantar2023gptq,
  author    = {Frantar, Elias and others},
  title     = {{OPTQ}: Accurate Post-Training Quantization for Generative Pre-Trained Transformers},
  booktitle = {ICLR}, year = {2023}
}

@inproceedings{dettmers2023qlora,
  author    = {Dettmers, Tim and others},
  title     = {{QLoRA}: Efficient Finetuning of Quantized {LLMs}},
  booktitle = {NeurIPS}, year = {2023}
}

@article{husom2025edge,
  author    = {Husom, Erik Johannes and others},
  title     = {Sustainable {LLM} Inference for Edge {AI}: Evaluating Quantized {LLMs} for Energy Efficiency, Output Accuracy, and Inference Latency},
  journal   = {ACM Transactions on Internet of Things}, year = {2025}
}

@article{kurt2026llamacpp,
  author    = {Kurt, Uygar},
  title     = {Which Quantization Should {I} Use? A Unified Evaluation of llama.cpp Quantization on {Llama-3.1-8B-Instruct}},
  journal   = {arXiv:2601.14277}, year = {2026}
}

@inproceedings{geiping2020inverting,
  author    = {Geiping, Jonas and others},
  title     = {Inverting Gradients: How Easy Is It to Break Privacy in Federated Learning?},
  booktitle = {NeurIPS}, year = {2020}
}

@inproceedings{schwarzschild2024acr,
  author    = {Schwarzschild, Avi and others},
  title     = {Rethinking {LLM} Memorization through the Lens of Adversarial Compression},
  booktitle = {NeurIPS}, year = {2024}
}

@article{shi2024muse,
  author    = {Shi, Weijia and others},
  title     = {{MUSE}: Machine Unlearning Six-Way Evaluation for Language Models},
  journal   = {arXiv:2407.06460}, year = {2024}
}

@article{clark2018arc,
  author    = {Clark, Peter and others},
  title     = {Think You Have Solved Question Answering? Try {ARC}, the {AI2} Reasoning Challenge},
  journal   = {arXiv:1803.05457}, year = {2018}
}

@inproceedings{zellers2019hellaswag,
  author    = {Zellers, Rowan and others},
  title     = {{HellaSwag}: Can a Machine Really Finish Your Sentence?},
  booktitle = {ACL}, year = {2019}
}

@article{sakaguchi2020winogrande,
  author    = {Sakaguchi, Keisuke and others},
  title     = {{WinoGrande}: An Adversarial {Winograd} Schema Challenge at Scale},
  journal   = {CACM}, year = {2021}
}
\end{document}